\documentclass[12pt]{iopart}

\usepackage{iopams}
\usepackage{graphicx}
\begin{document}

\title[]{Predictability of the Large Relaxations in a Cellular Automaton Model}

\author{Alejandro Tejedor}
\address{Faculty of Sciences, University of Zaragoza, Pedro
Cerbuna 12, 50009 Zaragoza, Spain.}

\author{Samuel Ambroj}
\address{Faculty of Sciences, University of Zaragoza, Pedro
Cerbuna 12, 50009 Zaragoza, Spain.}

\author{Javier B. G\'omez}
\address{Faculty of Sciences, University of Zaragoza, Pedro
Cerbuna 12, 50009 Zaragoza, Spain.}

\author{Amalio F. Pacheco}

\address{Faculty of Sciences, University of Zaragoza, Pedro
Cerbuna 12, 50009 Zaragoza, Spain.}

\begin{abstract}
A simple one-dimensional cellular automaton model with
threshold dynamics is introduced. It is loaded at a uniform rate and unloaded by abrupt relaxations. The cumulative distribution of the
size of the relaxations is analytically computed and behaves as a
power law with an exponent equal to $-1$. This coincides with the
phenomenological Gutenberg-Richter behavior observed in Seismology
for the cumulative statistics of earthquakes at the regional or
global scale. The key point of the model is the zero-load state of
the system after the occurrence of any relaxation, no matter what its
size. This leads to an equipartition of probability between all
possible load configurations in the system during the successive
loading cycles. Each cycle ends with the occurrence of the greatest
--or characteristic-- relaxation in the system. The duration of the
cycles in the model is statistically distributed with a coefficient
of variation ranging from 0.5 to 1. The predictability of the characteristic relaxations is evaluated by means of error diagrams.This model illustrates the value of taking into account the refractory periods to obtain a considerable gain in the quality of the predictions.

\end{abstract}

\pacs{89.75.Fb, 05.40.-a, 02.50.Ga}
\submitto{\JPA}
\maketitle 

\section{Introduction}\label{intro}
The cellular automaton model presented here will be denoted, for
brevity, as the C-Model (CM). It has analogies and differences with
the so called Minimalist Model (MM) \cite{MM1,MM2} that will be
emphasized in the next Sections. These types of cellular automaton models appeared in the context of the Self Organised Criticality (SOC) paradigm, introduced by Bak et al. \cite{Bak1,Bak2}, which meant a very important
new conceptual perspective to try to understand complexity in nature
in general and in natural hazards, such as forest-fires, landslides
and earthquakes, in particular. The flagship of  SOC is the
sand-pile model. It is a conservative cellular automaton where the size-frequency distribution of
avalanches exhibits a power-law behaviour  where the exponent corresponding to the non-cumulative distribution is around $-1$.

We will pay particular interest to
studying two properties of the CM, namely the probability of
occurrence of a relaxation of size $k$ for a system of size $N$, and
the probability distribution for the time interval between
successive characteristic relaxations, i.e. for the length of the
loading cycles. This distribution will be called $P_N(n)$, where $n$
is the discrete time elapsed since the occurrence of the last
characteristic relaxation. The mean and the variance of $P_N(n)$
will receive special consideration for the reasons that will be explained later.

For a better perspective of the results of this model we will refer to two important concepts coming from seismology. The best established law in regional seismicity is the Gutenberg-Richter relation between the magnitude of an earthquake
and its frequency. This law is of the power-law type when magnitudes
are expressed in terms of rupture areas
\begin{equation}\label{GR}
\dot N \propto S^{-b},
\end{equation}
where $\dot N$ is the number of observed earthquakes, per time unit,
with rupture area equal to or greater than $S$, and $b$ is the
so-called $b$-value, which is a universal phenomenological parameter
with a value close to unity \cite{Kagan1}. Note that equation
(\ref{GR}) represents a cumulative distribution. The corresponding
non-cumulative distribution would also be a fractal law with an
exponent equal to
\begin{equation}\label{b-value}
-(b+1)\sim -2.
\end{equation}
The interpretation of $b$ as a sort of universal critical exponent
in nature exactly equal to $1$ has spurred theoreticians for years
and many ideas, typically based on specific models and/or
mechanisms, have been offered to explain both the power law behavior
of (\ref{GR}) and the value of the $b$ parameter
\cite{VereJ,BakS,Ito,OFC,griegos,sotolongo}. Therefore, the Sand Pile model is not suitable for describing the observed Gutenberg-Richter Relation. But other SOC cellular automaton
models for earthquakes, such as those introduced by Olami, Feder and Christensen
\cite{OFC} provided, within a certain limit of conservation, a correct
order of magnitude for the $b$ parameter. For a review of SOC and
earthquakes, see \cite{Hergarten,Sornette}.

Besides, it is important to bear in mind that the above mentioned
Gutenberg-Richter law is a property of regional seismicity,
appearing when one averages seismicity over big enough areas and
long enough time intervals (e.g. \cite{Kossobokov}). For individual
faults, the type of size-frequency relationship is different from
the G-R law and consists of an approximate power-law distribution of
 small events (small compared with the maximum earthquake size
that a fault can support, given its area), which occur between
roughly quasi-periodic earthquakes of much larger size that rupture
the entire fault. These large quasi-periodic earthquakes are termed
``characteristic" \cite{S&C}, and the resulting size-frequency
relationship is the Characteristic Earthquake distribution. It must
be mentioned that the very concept of characteristic earthquake is
under debate \cite{Kagan2,Savage}.

Although the terminology that we use in this paper is often reminiscent of the seismological jargon, the CM could be applied to any entity where energy enters at a constant smooth rate and exits in the form of sudden relaxations with a very low efficiency because of dissipation.

The structure of the paper is as follows. In Section \ref{CMrules}, the
rules of the model are given together with a table and several
figures of numerical results. Section \ref{Analytic} is devoted to
some analytical deductions. In particular, using several properties
of Markov chains, we compute the stationary probabilities of the
configurations in the CM, the probability of occurrence of
relaxations of any size, and some other properties of the loading
cycle. Section \ref{forecasting} is devoted to the evaluation of the predictability of the large relaxations in this model. This is accomplished by using the so called error diagrams. A brief discussion and the conclusions are gathered in
Section \ref{Conclusions}. In a final Appendix we explain some
detailed calculations for the case $N=3$.

\section{\label{CMrules}The C-Model: rules and some numerical results}

The CM is a simple one-dimensional cellular automaton with threshold
dynamics. Consider a one-dimensional array of length $N$. The ordered positions in the
array will be labeled by an integer index $i$ varying from $1$ to
$N$. This system is loaded by receiving individual particles
in the various positions of the array, and unloaded by emitting
groups of particles through the first position $i=1$, which are
called relaxations or events. These two functions proceed using
the following five rules:

\begin{enumerate}
\item The incoming particles arrive at the system at a constant rate.
Thus the time interval between successive particles will be the
basic time unit in the evolution of the system. In comparison with
this time interval, the time taken by the relaxations is negligible.

\item All the $N$ positions have the same probability of receiving
a new particle. When a position receives a particle we say that it
is occupied.

\item The reception of a particle saturates a
position, i.e. if a new particle comes to a position already
occupied, that particle is dissipated.

\item The position $i=1$ is special. When a particle goes to the first
position a relaxation occurs. Then if all successive positions from
$i=1$ to $i=k$ are occupied and the position $k+1$ is empty, the
effect of the relaxation is to unload all the
levels from $i=1$ to $i=k$. Hence the size of the event is $k$.
The maximum relaxation corresponds to $k=N$, which is called the
characteristic relaxation of the system.

\item Besides, those positions that were occupied by particles but
were not affected by the relaxation lose their particles so
that the array is left completely empty.
\end{enumerate}

\begin{figure}
\begin{center}

\includegraphics[width= 12cm]{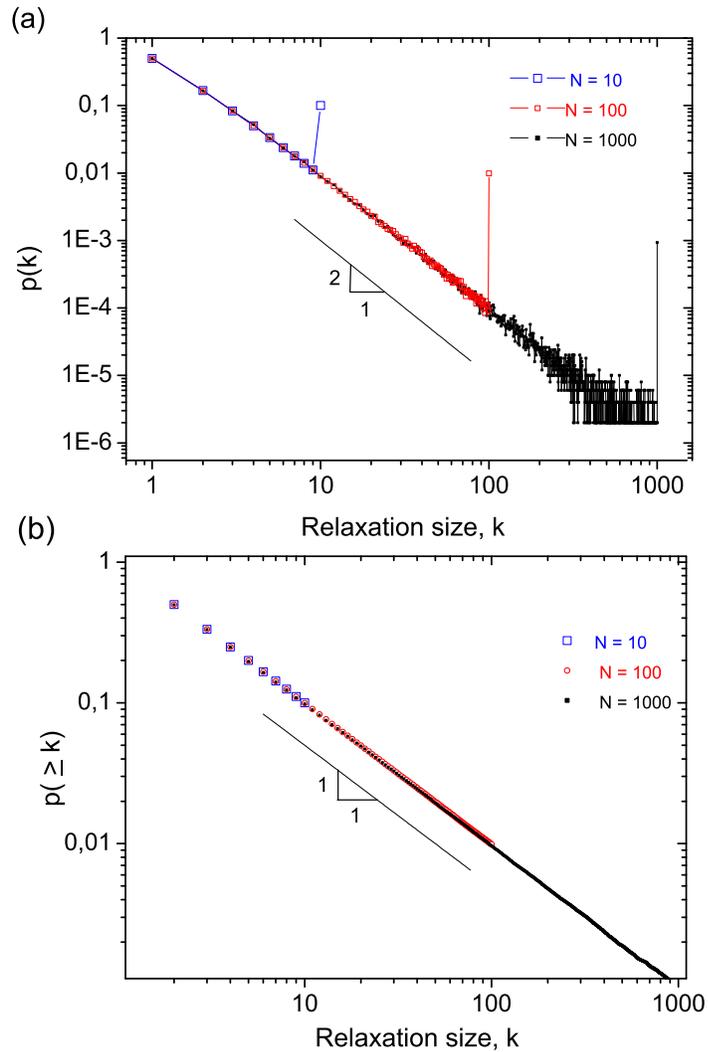} \caption{(a) Size-frequency plot for the
relaxations in the CM; the three cases $N=10$, 100, and 1000 are
superimposed. (b) cumulative magnitude-frequency plot. In both graphs
a line of slope $-2$ (top) and $-1$ (bottom) has been added for
visual comparison.} \label{Cmodel_GR}

\end{center}
\end{figure}

\begin{table}
\caption{Probability of occurrence of a relaxation
 of magnitude $k$ for a  system size, $N$.} \label{table}
\begin{center}
\begin{tabular}{lrrrrr}
\hline
\small{Size} & \small{N=2} & \small{N=3} & \small{N=4} & \small{N=10} & \small{N=100} \\
\hline
\small{k=1}   & \small{0.50006} & \small{0.50003} & \small{0.50000} & \small{0.50009} & \small{0.49948} \\
\small{k=2}   & \small{0.49994} & \small{0.16666} & \small{0.16652} & \small{0.16647} & \small{0.16667} \\
\small{k=3}   & -       & \small{0.33331} & \small{0.08333} & \small{0.08348} & \small{0.08359} \\
\small{k=4}   & -       & -       & \small{0.25015} & \small{0.04994} & \small{0.05012} \\
\small{k=5}   & -       & -       & -       & \small{0.03339} & \small{0.03358} \\
\small{k=6}   & -       & -       & -       & \small{0.02370} & \small{0.02370} \\
\small{k=7}   & -       & -       & -       & \small{0.01785} & \small{0.01768} \\
\small{k=8}   & -       & -       & -       & \small{0.01396} & \small{0.01413} \\
\small{k=9}   & -       & -       & -       & \small{0.01108} & \small{0.01118} \\
\small{k=10}  & -       & -       & -       & \small{0.10002} & \small{0.00906} \\
\small{k=99}  & -       & -       & -       & -       & \small{0.00009} \\
\small{k=100} & -       & -       & -       & -       & \small{0.00990} \\
\hline
\end{tabular}
\end{center}
\end{table}

Thus in the CM, after any small relaxation,
the system always restarts from the empty state, keeping no memory. The simplicity of these five rules  makes it very
appropriate for numerical simulations. The formalism and results of
the Markov Chains can also be applied to this model. In contrast with the CM, in the MM \cite{MM1}, rule 5 did not exist.
The fifth rule of this model could be representative of what occurs with the precipitation of a cloud. In the process of growth of the CCN (cloud condensation nuclei) by condensation, only those CCN that have been activated become cloud droplets and take part in the posterior processes of collision and coalescence, and therefore only they eventually contribute to the rain. The so-called haze droplets that were not activated evaporate.

\begin{figure}
\begin{center}

\includegraphics[width= 12cm]{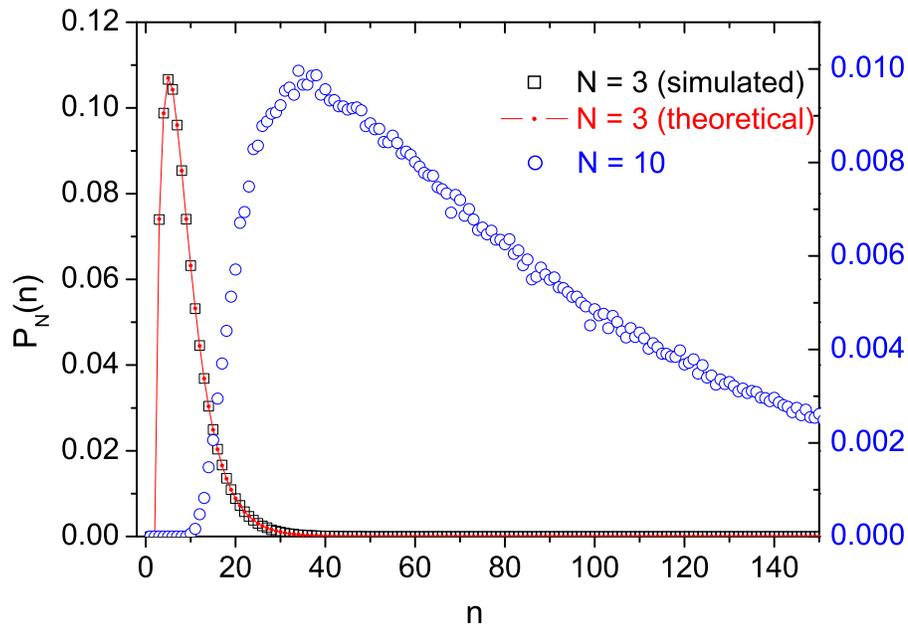} \caption{Probability distribution for the
length of the cycles, $P_N(n)$, for $N=3$ (squares, left axis), and
$N=10$ (circles, right axis). In the $N=3$ case the simulation
result is compared with the analytical result (continuous line). }
\label{Pn}

\end{center}
\end{figure}

Using these five rules and carrying out numerical simulations, we
have obtained the results contained in Table \ref{table} and in
figures \ref{Cmodel_GR}-\ref{alpha}, and figures \ref{ED_results} and
 \ref{N3Conf} later. From Table \ref{table} and Fig.\
\ref{Cmodel_GR}a we see that in the CM the value of $p_k(k<N)$ is an
$N$-independent constant. This sort of scale invariance, which also
appeared in the MM, can be justified by using the same arguments
written in Ref.\ \cite{MM1}. We also note that the non-cumulative
size-frequency plot for the events in this model is of the
Characteristic Earthquake type, as mentioned in the Introduction. In
Fig.\ \ref{Cmodel_GR}b, it is shown that, independently of $N$, the
cumulative size-frequency plot for the relaxations in this model is
a perfect power law with an exponent equal to $-1$. It is worth
saying at this point that, as far as the authors know, there is no
other model with this beautiful property. In the next Section, a
probabilistic argument will be provided for this notable result.

Note that the non-cumulative distribution in Fig.\ \ref{Cmodel_GR}a
clearly shows the special character of the maximum magnitude
relaxation, while the cumulative distribution hides this
preponderance. 

In Fig.\ \ref{Pn}, we show the probability distribution for the
length of the cycles, $P_N(n)$, for $N=3$ (squares) and $N=10$
(circles). In both cases the distribution shows an initial refractory period (i.e., an interval of null probability of relaxation
occurrence) of length equal to $N-1$, and a slow asymptotic decrease
which will be reflected in the high variances of this model. In case $P_3(n)$, the results obtained by
simulations are compared with the analytical result (continuous
line) obtained in the Appendix.

\section{\label{Analytic}Stationary Probabilities in the CM, Spectrum
of Earthquakes and other Analytic Properties}
In the CM, for a system of size $N$, there are $N$ groups of
configurations distinguished by their index of load occupation
$\theta$,
\begin{equation}\label{theta}
\theta = 0,1,2,\dots,N-1.
\end{equation}
In each of these groups there are
\begin{equation}\label{theta_bin}
\left( {\begin{array}{*{20}c} N-1 \\ \theta \\ \end{array}} \right)
\end{equation}
different stationary configurations denoted by $\theta_j$, where the
integer index $j$ runs in the range
\begin{equation}\label{theta_range}
1 \leq j \leq \left( {\begin{array}{*{20}c} N-1 \\ \theta \\
\end{array}} \right).
\end{equation}
As in any Markov chain of this type, if the Markov matrix is denoted
by $\mathbf{M}$, then the stationary probabilities for each
configuration are the components of the eigenvector of
$\mathbf{M}^{T}$ corresponding to the eigenvalue $+1$. These
components will be denoted by $\pi(\theta_j)$. Their $j$-independent
values are
\begin{equation}\label{pi_theta}
\pi(\theta_j)= \frac{1}{N} \frac{1}{\left( {\begin{array}{*{20}c} N-1 \\ \theta \\
\end{array}} \right)}.
\end{equation}
Relations (\ref{pi_theta}) are identified in a straightforward way
because of the regularity of the Markov matrix $\mathbf{M}$ in this
model (as an illustration, the case $N=3$ is worked out in detail in
the Appendix). Thus, all the configurations that share the same
$\theta$ have the same probability, and the sum of the probabilities
of all of them is $1/N$. In other words, any load $\theta$ in this
system has identical probability $1/N$, and the probability among
all the possible configurations corresponding to a given $\theta$
is also equally distributed.

\begin{figure}
\begin{center}

\includegraphics[width= 12cm]{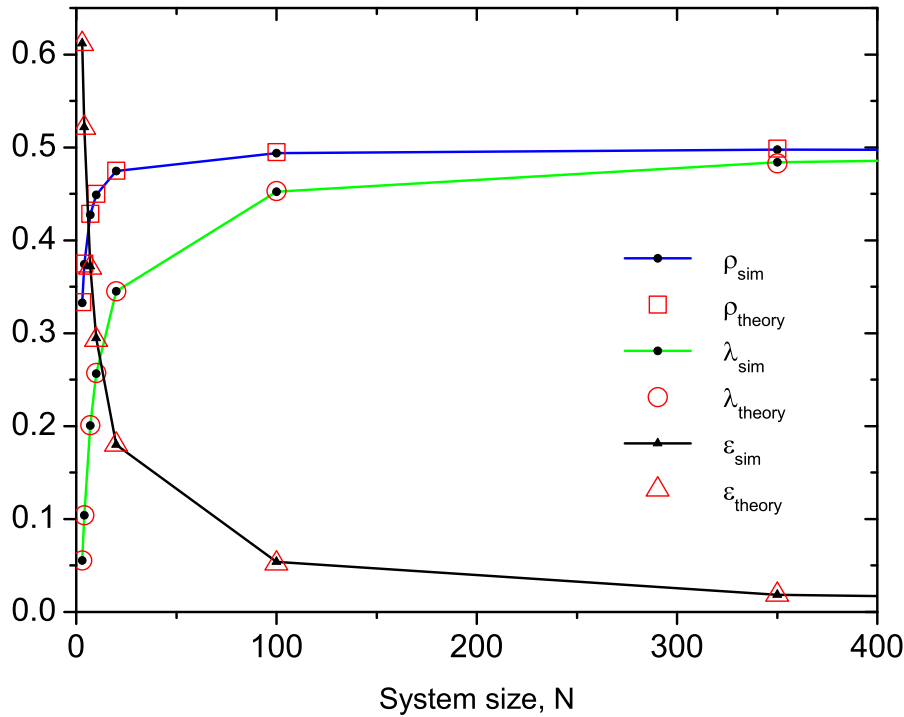} \caption{ Reflectivity $\rho$, efficiency $\epsilon$, and fraction of lost particles during relaxations
$\lambda$ as a function of $N$. For details, see Section
\ref{Analytic}. } \label{diss}

\end{center}
\end{figure}

Using these results, let us compute first the fraction of particles
that, on their way to a position in the array, find this position
occupied and are therefore lost. This proportion will be called the
reflectivity $\rho$ of the system:
\begin{eqnarray}\label{rho}
\rho &=& \sum_{\theta_j}\pi(\theta_j) \frac{\theta}{N} = \sum_\theta
\left( {\begin{array}{*{20}c} N-1 \\ \theta \\
\end{array}} \right) \frac{1}{N} \frac{1}{\left( {\begin{array}{*{20}c} N-1 \\ \theta \\
\end{array}} \right)} \frac{\theta}{N} \nonumber \\
&=& \frac{1}{N^2}\sum_{\theta=0}^{\theta=N-1}\theta =
\frac{1}{N^2}\frac{N(N-1)}{2} = \frac{1}{2}
\left(1-\frac{1}{N}\right).
\end{eqnarray}
Thus, for large $N$ nearly 50\% of the particles are lost by
reflection. This result is  compared with the simulations in figure
\ref{diss}. We see there that the reflectivity (open and filled
squares) increases with the system size, abruptly at the beginning and
more slowly for bigger system sizes, reaching an asymptotic value of
$\sim 0.5$ for big systems ($N \geq 200$). The smallest reflectivity
is $\rho \sim 0.34$ for $N=3$.

\begin{figure}
\begin{center}

\includegraphics[width= 12cm]{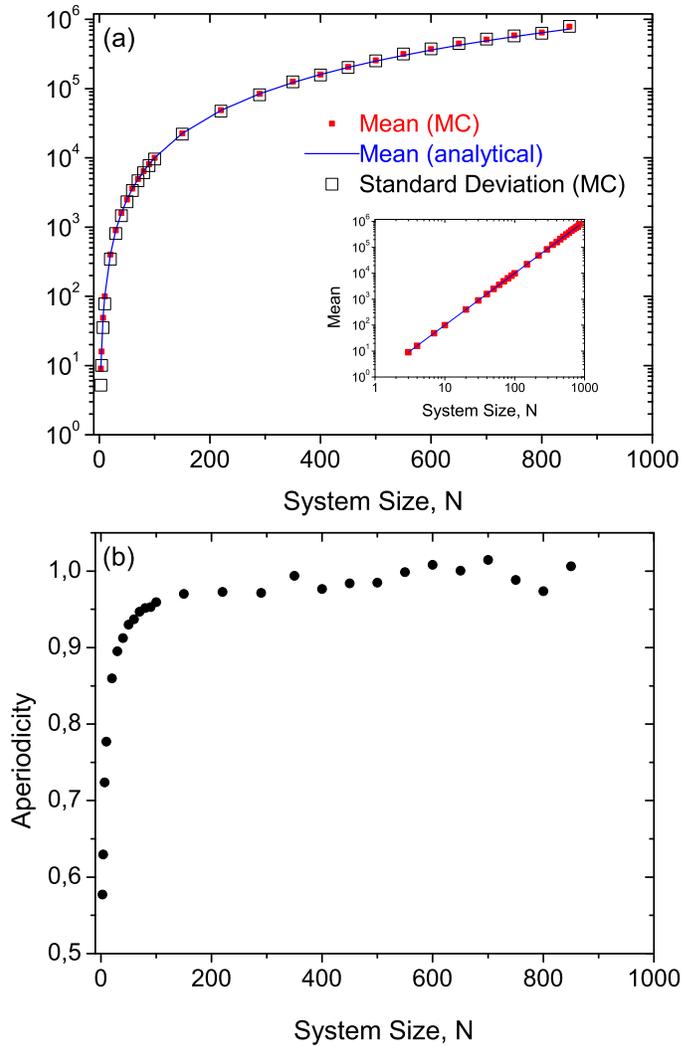} \caption{ (a) Mean and standard deviation of
$P_N(n)$ as a function of $N$; the inset shows the trend of the mean
in log-log scales. MC stands for MonteCarlo results. (b)
Aperiodicity versus $N$. } \label{alpha}

\end{center}
\end{figure}

A second analytical conclusion is the following. From Eq.(\ref{pi_theta}), we know that in the case
of maximum occupation, i.e., when there are $N-1$ particles in the
system, its probability is
\begin{equation}\label{pi}
\pi(N-1)=\frac{1}{N},
\end{equation}
therefore the probability that at any arbitrary time a
characteristic event occurs is $1/N^2$ and, in consequence, the
mean value of the time elapsed between two consecutive
characteristic relaxations in this model is
\begin{equation}\label{mean}
<P_N(n)>=N^2.
\end{equation}
This conclusion coincides with the numerical result presented in
figure \ref{alpha}a, better seen in the inset where log-log scales
have been used to appreciate the +2-slope trend of the mean. Note
also how, for small system sizes, the mean is bigger than the
standard deviation but that both of them tend to a common value
as the system size increases. This can be better appreciated in figure
\ref{alpha}b where the aperiodicity, i.e. the ratio of the standard
deviation to the mean, is plotted. The minimum
aperiodicity is 0.57 ($N=3$) and grows asymptotically to unity.

We also know that the mean time between two arbitrary consecutive
relaxations in the model is $N$, and therefore we deduce that in
this model, between two consecutive characteristic events,
there are, on average, $N-1$ non-characteristic relaxations.

As mentioned above, a cycle is the time between two successive
characteristic events. And each cycle is formed by a succession
of several initial sub-cycles, marked by the occurrence of non
characteristic events, and a {\it final} sub-cycle marked by
the occurrence of the characteristic relaxation that ends the cycle.
The calculation of the length of the final sub-cycle is
straightforward. This time length is formed by the addition of two
independent stages. The first stage, which starts from the state
$\theta = 0$ and ends when the system arrives to the state $\theta =
N -1$, will be called the {\it loading stage}. The second, in which
the system is completely loaded, ends when a particle hits the first
site; this will be called the {\it hitting stage}. The length of the
loading stage is the mean time taken to fill a box of $N -1$ cells
by random assignments of the particles taking into account the
reflection of particles by occupied cells \cite{BoxModel}. The
length of the hitting stage is the mean time of a geometric
distribution with a hitting probability equal to $1/N$. Thus, we
have
\begin{equation}\label{subcycle}
<n>_{fsc}=<n>_{l}+<n>_{h}=N(S-1)+N,
\end{equation}
where $S$ is a function of $N$ defined as
\begin{equation}\label{S}
S=\sum_{k=1}^{N}\frac{1}{k}.
\end{equation}
For $N\geq 10$, $S$ is already well approximated by
\begin{equation}\label{limS}
\lim_{N\rightarrow\infty}S =C+\ln N
+\frac{1}{2N}+\mathcal{O}\left(\frac{1}{N^2}\right),
\end{equation}
where $C = 0.577215\dots$ is Euler's constant. Therefore, the
asymptotic behavior of the duration of the final sub-cycle is of the
form
\begin{equation}\label{limDur}
\lim_{N\rightarrow\infty}<n>_{fsc} =N\ln N.
\end{equation}

The variance of the time length of the final sub-cycle can  also be analytically calculated.

Using numerical simulations, we have also seen in Section
\ref{CMrules} that the CM provides a power-law relation, with $b=1$,
for the cumulative size-frequency distribution of the relaxations.
This power-law relation can be understood by a simple probabilistic
argument. In the CM, after any relaxation, the grid contains no
particles, i.e., all the positions of the system are empty.
Therefore, to have in the next event a relaxation of size $\geq k$
$(1\leq k \leq N)$, it is necessary that the process of occupation
of the lowest $k$ levels be in such a way that the first level
($i=1$) is the last one to be occupied. The question then is: What
is the probability that $k$ identical sites be occupied with the
condition that one of them be the last? As all of the sites are
equivalent, the answer is
\begin{equation}\label{last}
\frac{1}{k}.
\end{equation}
Thus, for $1\leq k \leq N$,
\begin{equation}\label{Paccum}
p(\geq k) = \frac{1}{k}
\end{equation}
independently of the size $N$ of the system. 

Now, due to the fact that
\begin{equation}\label{Pdif1}
p(k) = p(\geq k) - p(\geq k+1),
\end{equation}
which can be applied for $1 \leq k \leq N-1$ , we find that
\begin{eqnarray}\label{Pdif2}
p(k) = \frac{1}{k}-\frac{1}{k+1} = \frac{1}{k(k+1)} =
\frac{1}{k^2(1+1/k)},\nonumber \\
k\leq N-1.
\end{eqnarray}
For the limit case $k=N$, we apply expression (\ref{Paccum}) and
obtain
\begin{equation}\label{Pdif3}
p(k=N) = p(k\geq N) = \frac{1}{N}.
\end{equation}
The verification of the adequate normalization of $p(k)$ is carried
out as follows:
\begin{eqnarray}\label{norm}
\sum_{k=1}^{N-1}p(k)+p(N) &=& \sum_{k=1}^{N-1}\left( \frac{1}{k} -
\frac{1}{k+1}\right)+\frac{1}{N} \nonumber \\
&=& 1-\frac{1}{N}+\frac{1}{N}=1.
\end{eqnarray}
Therefore, in the CM the cumulative spectrum of relaxation given in
(\ref{Paccum}) constitutes a power law with an exponent $-1$. This
coincides with the phenomenological Gutenberg - Richter exponent
observed in seismology for the cumulative statistics at the regional
or global scale. And according to this, in the CM the non-cumulative
form of the size-frequency relation, as given in (\ref{Pdif2}), is
asymptotically a power law with an exponent equal to $-2$. These
conclusions exactly reproduce the plots in figure \ref{Cmodel_GR}.
Relation (\ref{Pdif1}) is the discrete derivation of (\ref{Paccum});
in the continuum it would correspond to
\begin{equation}\label{Pcont}
-\frac{d}{dx}\frac{1}{x} = \frac{1}{x^2}.
\end{equation}

With respect to the probability partition that relates a system of
size $N$ with the next one of size $N+1$, that is, the form in which
relation
\begin{equation}\label{Prec1}
p_N(N) = p_{N+1}(N) + p_{N+1}(N+1)
\end{equation}
is implemented in this model, we have the following simple
form
\begin{equation}\label{Prec2}
\frac{1}{N} = \frac{1}{N(N+1)} + \frac{1}{N+1}.
\end{equation}
Here, it is interesting to note that the counterpart of equation
(\ref{Prec2}) in the MM can only be obtained numerically.

Let us now compute the efficiency of the system $\epsilon$, i.e., the
proportion of particles that contribute to the relaxations of the
system. We know that, on average, in $N$ time units a relaxation is
produced. Thus,
\begin{equation}\label{epsilon1}
\epsilon=\frac{1}{N}\left[\sum_{k=1}^{N-1}
\frac{1}{k^2\left(1+\frac{1}{k}\right)}k+\frac{1}{N}N\right].
\end{equation}
And using the function $S(N)$ as defined in (\ref{S}), the
efficiency can be expressed as
\begin{equation}\label{epsilon2}
\epsilon=\frac{S}{N}.
\end{equation}
This function amounts to $\epsilon= 0.61$ for $N=3$ and monotonously
decays to $0$ as $N$ grows (figure \ref{diss}, open and filled
stars). Thus, we realise that this model is by far less generous than a standard slot machine. For example, present laws in Nevada impose a minimum of  $\epsilon=0.75$ for the payout percentage of slot machines in the casinos of this state. For large $N$,
\begin{equation}\label{limEps}
\lim_{N\rightarrow\infty}\epsilon = \frac{\ln N}{N}.
\end{equation}
In this stochastic model, the history of one particle can be only
one of the three following things: (i) reflected by the system, (ii)
emitted in a relaxation, or (iii) lost on the occasion of a relaxation in the system. Calling $\lambda$ to the fraction of these
lost particles, we have
\begin{equation}\label{lambda}
\lambda = 1-\rho - \epsilon.
\end{equation}
Due to the fact that $\epsilon \rightarrow 0$ for large $N$, in this
limit both $\rho$ and $\lambda$ share one half of the probability.
This is illustrated in figure \ref{diss} where we see how $\lambda$
(open and filled circles) grows from $\lambda \sim 0.05$ for $N=3$
to $\lambda=0.5$ for big system sizes.

\section{\label{forecasting}Forecasting the Large-Avalanche Occurrence.}
\begin{figure}
\begin{center}
\noindent\includegraphics[width=8cm]{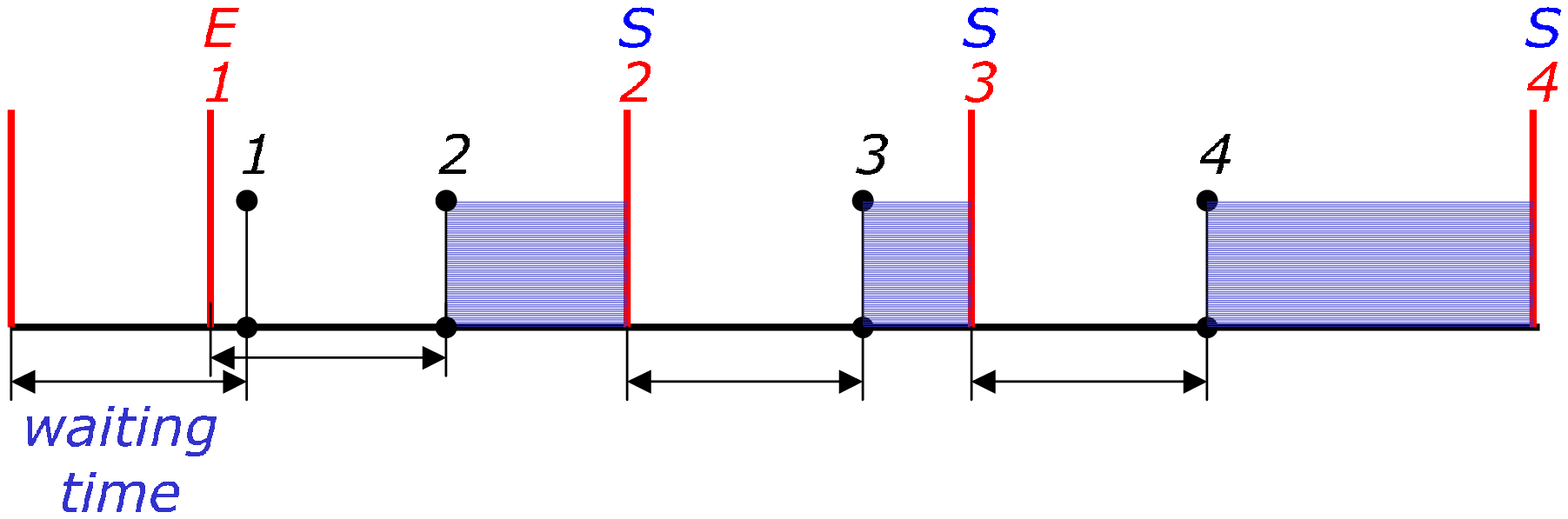}

\noindent\includegraphics[width=4cm]{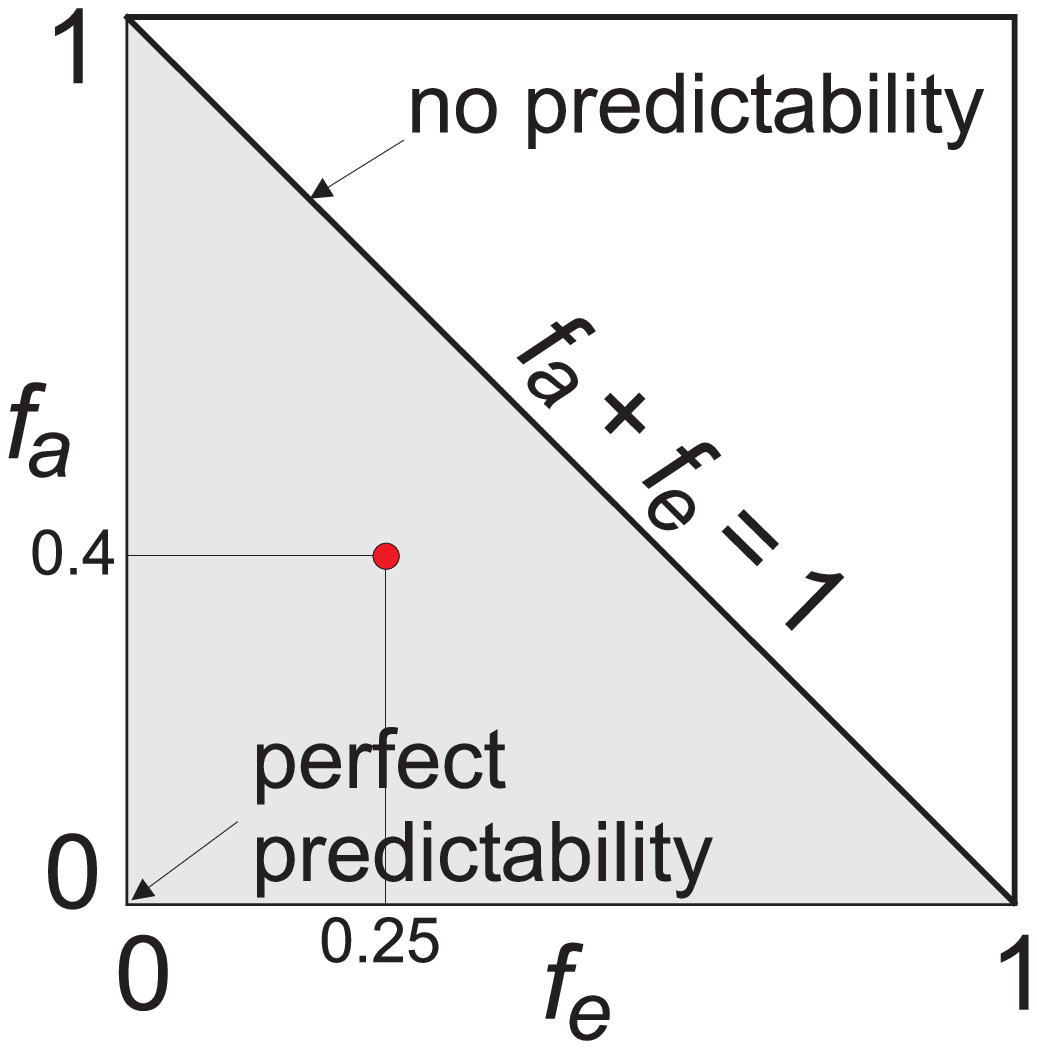}\caption{\label{ED_def}
Reference strategy for the assessment of the predictability of a time series and its representation on an error diagram. The events that are to be predicted (large relaxations) are the vertical red bars numbered correlatively. An alarm is set a fixed time interval after each event (waiting time) and the prediction is labeled error (E) or success (S) depending on whether the alarm was off or on when the event occurred. The fraction of errors is the number of events not predicted (one in the example) divided by the total number of events (four events), i.e., $f_e=0.25$; and the fraction of alarm time is the total alarm time (blue sections of the time line, 71 time units) divided by the total duration of the time series(178 time units), i.e., $f_a=0.4$ in the example shown in the figure.}
\end{center}
\end{figure}

A hint of the predictability of the large relaxations in this type of model is given by the aperiodicity of their time series. The aperiodicity is a quantitative measure of the lack of regularity of a time series. If $\mu$ is the average time
between two consecutive characteristic relaxations (i.e., the mean
duration of the cycle), and $\sigma$ is the standard
deviation of the duration around the mean, then $\alpha = \sigma /
\mu$. The aperiodicity is otherwise known as the coefficient of
variation. A value of $\alpha =0$ means that the system is {\it
periodic} and all the cycles have exactly the same duration;
the range $0<\alpha<1$ defines {\it quasi-periodic} time series; and the case $\alpha=1$ can correspond to a purely random (Poissonian) time series, but not necessarily so. Time series with $\alpha>1$
are said to be {\it clustered}.
As the $\alpha$ of this model takes values between $0.5$ and $1$, the occurrence of the large events is a quasi-periodic phenomenon.
A robust way to assess the predictability of a time series is by trying to forecast its events by declaring {\em alarms} at particular times (Figure \ref{ED_def}). The aim is to declare alarms before {\em all} the events in order not to miss any event, but to declare them {\em just} before the events in order to minimize the total alarm time. Many strategies can be devised to declare the alarms but there is a reference strategy to which all others can be compared \cite{MM2,NewmanTurcotte02,KeilisBorok03}. This strategy consists of setting the alarm a fixed time interval after each event (waiting time, $t$) and maintaining it until the occurrence of the event. If the following event in the time series occurs before the alarm is raised, it is counted as a prediction error; if the following event in the time series occurs after the alarm is raised, it is counted as a prediction success and the alarm is then cancelled. 

The fraction of errors $f_e$ (number of missed events divided by the total number of events) and the fraction of alarm time $f_a$ (total alarm time divided by the total duration of the time series) can be computed as a function of the above mentioned waiting time $t$, and the purpose is to find the optimum waiting time. This optimum waiting time depends on the relative importance that failing to predict an event has compared to keeping the alarm on. An objective function, called \textit{loss function}, $L$, can be defined that incorporates this trade-off in each particular case. Here we will use the simplest of them, $L = f_e + f_a$, where failure to predict and a long alarm time are equally penalized. Thus, our aim is to find the waiting time $t=t^\star$ that minimizes $L(t)$. This minimum value is denoted by $L^\star = L(t^\star)$. And the best way to graphically display this is by means of an error diagram, where the fraction of errors $f_e$ runs along the horizontal axis and the fraction of alarm time $f_a$ runs along the vertical axis (Fig.\ \ref{ED_def}). Error diagrams were introduced in earthquake forecasting
by Molchan \cite{Molchan97} who contributed with rigorous mathematical
analysis to the optimization of the earthquake prediction
strategies.
\begin{figure}
\begin{center}

\noindent\includegraphics[width=12cm]{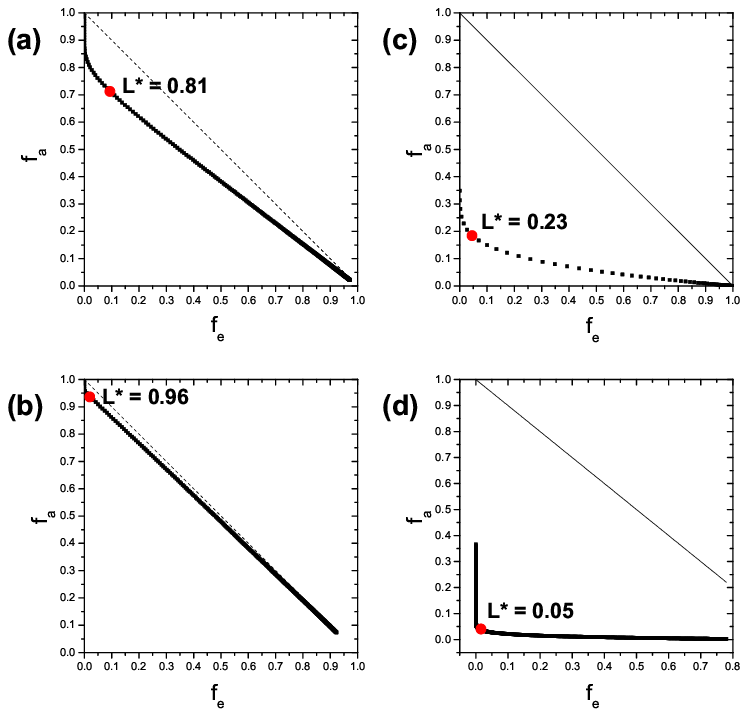}
\caption{\label{ED_results}Error Diagrams (a)(c) for  $N=10$ and (b)(d) for $N=100$.  The results of the Reference Strategy are shown at the left. The right figures show the improvement in the predictability provoked by considering the refractory periods. For details see the text.  }

\end{center}
\end{figure}

Figures \ref{ED_results}a y \ref{ED_results}b show the results of applying the reference strategy to a system of $N=10$ and $N=100$ . For $N=10$ : $t^*=29$, $f_e=0.09$, $f_a=0.72$ and $L^*=0.81$; for $N=100$ : $t^*=588$, $f_e=0.02$, $f_a=0.94$ and $L^*=0.96$.  This increase in $L^*$ is what one would reasonably expect as the aperodicity of these systems is $\alpha=0.78$ for $N=10$ and $\alpha=0.96$ for $N=100$. The question now is, can we perform  a better job in forecasting the occurrence of the large relaxations in this model? The answer is definitely yes. For this goal we will exploit the conspicuous refractory periods  that take place after the occurrence of any relaxation in the model. 

The reference strategy consists in connecting the alarm at $t$ time units after the occurrence of each characteristic relaxation in the system, and the best $t$ is that which minimizes the loss function $L$.  As the alarm is not disconnected until the occurrence of the next characteristic relaxation, in this strategy there are no false alarms. 
The new strategy consists in connecting the alarm at $(N-1)+t'$ time units after the occurrence of any relaxation, big or small, in the system. And the best $t'$ is that which minimizes L.  As in this strategy the alarm is disconnected when any relaxation takes place, we will have false alarms that will contribute to the total $f_a$.  The results of the new strategy are plotted in figures \ref{ED_results}c and \ref{ED_results}d.  The best values of the parameters for $N=10$ are ${t'}^*=7$, $f_e=0.05$, $f_a=0.18$ and $L^*=0.23$; and for $N=100$ are ${t'}^*=214$, $f_e=0.01$, $f_a=0.04$ and $L^*=0.05$. It is clear that the new strategy has been a success.

\section{\label{Conclusions}Discussion and Conclusions}
The CM presented here is a new model of the type that started
with the MM \cite{MM1,MM2}. In these models, the only parameter to
be fixed is $N$, the total number of sites, or cells,
susceptible to occupation. The two models are materialized on
discrete one-dimensional arrays, which permits their study by means
of Markov chains or straightforward Montecarlo simulations. 

The CM is by no means conservative: the load is dissipated in
several forms. The array that is loaded and unloaded in
successive cycles dissipates load by reflection of the already
occupied sites, and when relaxations are produced. As a result of
its rules, this system maintains a statistical equilibrium such that
all possible values of the load in the system have the same probability. Besides,
all the configurations with the same load are also equally probable.
The non-cumulative size-frequency relation for all the relaxations
in this model is of the Characteristic-Earthquake type, with an
excess of system-wide events. After an initial transient, the
probability of the small relaxations is a power law with an exponent
equal to $-2$ (see figure \ref{Cmodel_GR}a). In consequence, the
cumulative size-frequency relation for all the events in this
model is a neat power-law with exponent $-1$. This is notable
because it coincides with the form of the Gutenberg-Richter law of
the global seismicity.

The lack of memory in the CM with respect to the MM makes this model
simpler. A clear manifestation of this simplicity can be observed in
the computation of the probability distribution of the length of the
cycles when the Jordan form of the respective $\mathbf{M}'$
matrices are computed in the CM (see the Appendix for the definition
of matrix $\mathbf{M}'$ and other algebraic details). In the CM the
Jordan form of $\mathbf{M}'$ is a diagonal matrix and therefore
$P_N(n)$ is just a combination of exponentials (with $n$ in the
exponents). In the MM, however, the appearance of off-diagonal terms
in $\mathbf{M}'$ led $P_N(n)$ to be formed by terms where polynomial
terms in $n$ are multiplied by exponential terms in $n$
\cite{GomezP}. 

In spite of the fact that in the CM the period of return of the big relaxations has a aperiodicity between 0.5, for small N, and tends to 1, for large N, we have shown that these big relaxations can be accurately predicted by exploiting the refractory periods that take place after the occurrence of any relaxations in the system. In fact, it has been shown that, the bigger N is, the better can be the result of the forecasting. 

There are plenty of natural systems where just after the occurrence of a characteristic event, the probability of occurrence of a new one is null or very low. Examples of this type of inhibition are the process of firing of neurons, the refractory periods that occur between the breakouts of contagious diseases, etc. This model has illustrated the importance of taking into account these periods for making accurate predictions.

\subsection{Note Added.}
An anonimous referee has suggested us to make a comparison between the model studied in this work, the CM (and especially the MM) and those models recently used to describe the dynamics of microtubules. 
Microtubules are linear polymers that serve as structural components within cells and are involved in many cellular processes including vesicular transport. In its simplest version, called the minimalist model \cite{ANTAL} microtubules are represented by a chain formed by two types of monomers: GTP (for guanosine triphosphate) and GDP (for guanosine diphosphate). The state of a microtubule evolves due to the three following three processes:

\begin{enumerate}
\item Attachment.  A microtubule grows by attachment of a GTP monomer at its tip, with a rate $\lambda$  if in the tip there is a monomer GTP, and with a rate $p\lambda$  if in the tip there is a GDP monomer
\item Conversion. Each GTP monomer of the chain can be independently converted by hydrolysis into a GDP monomer. This occurs at a rate unity.
\item Detachment. A microtubule shrinks due to the detachment of a GDP monomer that is situated at the tip of the microtubule. This process takes place at a rate $\mu$ . 
\end{enumerate}

In the phase space subtended by the set of parameters ($\lambda$, $\mu$, p) there is a rich phenomenology for the microtubule growth dynamics such as phase boundaries separating regions where the microtubule grows, on average, at a certain rate from other regions where the average length of the microtubule is finite, etc.  And as a glaring event in the limit of large $\mu$ it is worth mentioning that a global catastrophe can occur when a newly attached GTP at the tip converts to a GDP and the rest of the microtubule consists only of GDP at that moment, the microtubule instantaneously shrinks to zero length.
\\

And the question now is which are the differences and similarities between these idealized microtubule growth models (MGM) and our CM and MM? Both type of models are stochastic and in one spatial dimension but while the MGM describe a  system where its length can grow or shrink, i.e. fluctuates in time,  in both the CM and the MM the length of the array where the particles are positioned, $N$, is fixed. Obviously this difference is crucial. What fluctuates in our models is the number of particles, $k$, emitted in a relaxation, which varies between 1 and $N$. 
\\

Having stated the main difference, the question would be: could we modify our models to resemble the MGM? Maintaining all the other rules of these models, this task would require addding new basic ingredients. Assuming that the first site corresponds to the tip of the polymer, a new parameter $\alpha$ with dimensions of a rate would be responsible of adding new free sites at the tip. Second, after the occurrence of a relaxation of $k$ particles, the $k$ sites of the array neighbour of the tip would disappear and the $k+1$  position of the array, which is free, would turn into the new first site responsible of controlling the new relaxation. Thus, in a short dictionary between the two models a GTP monomer would be represented by a free site, a GDP monomer by an occupied  site, the reception of a particle shifts the site from GTP to GDP, a relaxation shrinks the length, and characteristic relaxation would be the equivalent to the catastrophe of the microtubule. The new MG-like model sketched here would be more economical in parameters than those studied in \cite{ANTAL}.  Its properties will be studied in a next future.

\appendix
\section{Explicit calculations for $N=3$}

In this Appendix the algebraic calculations corresponding to the
case $N = 3$ are worked out in detail. These results will likely be
useful to the reader for a better understanding of the content of
Section \ref{Analytic}.

\begin{figure}
\begin{center}

\includegraphics[width= 12cm]{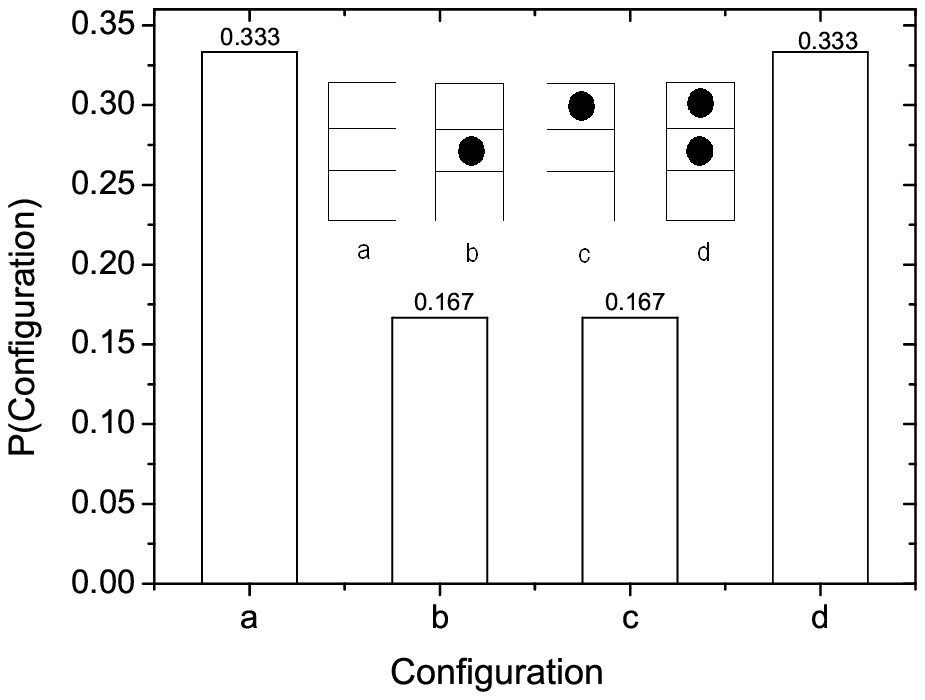} \caption{ Configurations and residence
probabilities for the case $N=3$. Only four states, labeled a, b, c,
and d, are possible (inset). The probability of states a and d is
1/3 and the probability of states b and c is 1/6. } \label{N3Conf}

\end{center}
\end{figure}

A scheme of the possible states for this system is plotted in Fig.\
\ref{N3Conf}. Note that the relation between this notation and the
more general notation expressed in equations (\ref{theta}),
(\ref{theta_bin}) and (\ref{theta_range}) is $a\equiv(\theta=0)$,
$b\equiv(\theta=1,\;j=1)$, $c\equiv(\theta=1,\;j=2)$,
$d\equiv(\theta=2)$.

Using the rules of the model, the probabilities of the non-null
one-step transitions are: $\mathbf{M}_{a\rightarrow a}=1/3$,
$\mathbf{M}_{a\rightarrow b}=1/3$, $\mathbf{M}_{a\rightarrow
c}=1/3$, $\mathbf{M}_{b\rightarrow a}=1/3$,
$\mathbf{M}_{b\rightarrow b}=1/3$, $\mathbf{M}_{b\rightarrow
d}=1/3$, $\mathbf{M}_{c\rightarrow a}=1/3$,
$\mathbf{M}_{c\rightarrow c}=1/3$, $\mathbf{M}_{c\rightarrow
d}=1/3$, $\mathbf{M}_{d\rightarrow a}=1/3$,
$\mathbf{M}_{d\rightarrow d}=2/3$. Thus, for $N=3$, the Markov
matrix is
\begin{equation}\label{M}
\mathbf{M}=
\left(%
\begin{array}{cccc}
  1/3 & 1/3 & 1/3 & 0 \\
  1/3 & 1/3 & 0 & 1/3 \\
  1/3 & 0 & 1/3 & 1/3 \\
  1/3 & 0 & 0 & 2/3 \\
\end{array}%
\right).
\end{equation}
The stationary probabilities of residence in this Markov system are
the normalized components of the eigenvector of $\mathbf{M}^T$
corresponding to the eigenvalue $\lambda = 1$, $\mathbf{M}^T$ being
the transpose of $\mathbf{M}$. These components are
\begin{equation}\label{eigen}
\left| \lambda=1 \right\rangle = \frac{1}{3}
\left(%
\begin{array}{c}
  1 \\
  1/2 \\
  1/2 \\
  1 \\
\end{array}%
\right),
\end{equation}
which agrees with that written in (\ref{theta_bin}) and
(\ref{pi_theta}) for any $N$. In the calculation of the probability
distribution of the length of the cycles we will follow the general
method used in \cite{MM1}, that is
\begin{equation}\label{PnA}
P_N(n)=\frac{1}{N}[\mathbf{M}']^{n-1}(\theta=0,\,\theta=N-1).
\end{equation}
This equations means that the probability of having a cycle of
length $n$ is given by $1/N$ times element ($\theta=0,\,\theta=N-1$)
of the matrix $[\mathbf{M}']^{n-1}$. This is justified as follows.
On the left, we have, by definition,
 the probability that the cycle lasts $n$ steps. On the right, it is
 written the probability that, in $n-1$ steps, the system has moved
 from the empty state to the state with a load $N-1$, with the
 restriction that this is the first visit to this  maximally
 loaded state. This restriction is easily implemented by making the matrix
 element $(\theta=N-1,\;\theta=0)$  in $\mathbf{M}$ equal to $0$.
 That ``pruning'' converts matrix $\mathbf{M}$ into the
 new matrix $\mathbf{M}'$. Once the system has done this
 transit in $n-1$ steps, with the mentioned restriction, the probability of passing
 in one step from the state $\theta=N-1$ to the state $\theta=0$ is $1/N$.
 This is the reason to include this factor in (\ref{PnA}).

 In the case $N=2$ the MM \cite{MM1} and the CM are actually the same model. The
 $P(n)$ distribution is 
 \begin{equation}\label{Pn2}
P_{N=2}(n) = \frac{n-1}{2^n}.
\end{equation}
In the case $N=3$, $\mathbf{M}'$ is
\begin{equation}\label{Mprime}
\mathbf{M}'=\frac{1}{3}
\left(%
\begin{array}{cccc}
  1 & 1 & 1 & 0 \\
  1 & 1 & 0 & 1 \\
  1 & 0 & 1 & 1 \\
  0 & 0 & 0 & 2 \\
\end{array}%
\right).
\end{equation}

In order to obtain a generic power of this matrix, we will perform
its Jordan decomposition
\begin{equation}\label{Jordan}
\mathbf{M}'=\mathbf{QJQ}^{-1}
\end{equation}
and hence
\begin{equation}\label{Jordan_power}
\mathbf{M}'^{n-1}=\mathbf{QJ}^{n-1}\mathbf{Q}^{-1},
\end{equation}
where the factors on the right of (\ref{Jordan_power}) are
\begin{eqnarray}\label{factors}
\mathbf{Q} &=&
\left(%
\begin{array}{cccc}
  0  & -2 & -\sqrt{2} & \sqrt{2} \\
  -1 & -1 & 1         & 1 \\
  1  & -1 & 1         & 1 \\
  0  & 1  & 0         & 0 \\
\end{array}%
\right), \nonumber \\
\mathbf{J}^{n-1} &=& \left(\frac{1}{3}\right)^{n-1} \times \nonumber \\
& &
\left(%
\begin{array}{cccc}
  1 & 0       & 0                  & 0 \\
  0 & 2^{n-1} & 0 & 0 \\
  0 & 0       & (1-\sqrt{2})^{n-1}                  & 0 \\
  0 & 0       & 0                  & (1+\sqrt{2})^{n-1} \\
\end{array}%
\right), \nonumber \\
\mathbf{Q}^{-1} &=&
\left(%
\begin{array}{cccc}
  0                    & -1/2 & 1/2 & 0 \\
  0                    & 0    & 0   & 1 \\
  -\frac{1}{2\sqrt{2}} & 1/4  & 1/4 & -\frac{4-2\sqrt{2}}{4\sqrt{2}} \\
  \frac{1}{2\sqrt{2}}  & 1/4  & 1/4 & -\frac{-4-2\sqrt{2}}{4\sqrt{2}} \\
\end{array}%
\right). \nonumber \\
\end{eqnarray}

Inserting the corresponding formulae into (\ref{PnA}), we obtain
\begin{eqnarray}\label{PnN3}
P_{N=3}(n) = \left(\frac{1}{3}\right)^n
\left\{-2^n+\frac{2-\sqrt{2}}{2(1-\sqrt{2})}(1-\sqrt{2})^n+ \right. \\
\nonumber \left. \frac{2+\sqrt{2}}{2(1+\sqrt{2})}(1+\sqrt{2})^n
\right\}, n\geq 2.
\end{eqnarray}
This function has been plotted in figure \ref{Pn} together with its
numerical counterpart. It is null for $n = 2$, as it should, because
for $N = 3$ the minimum length of a cycle is $3$. It is apparent
that $P_{N=3}(n)$ is formed by a combination of three geometric-like
terms. This is due to the fact that matrix $\mathbf{J}$ has no
off-diagonal term. Inside the brackets the first term is negative,
the second is oscillating, and the third --and dominant-- is
positive. Therefore, in the limit of large $n$ this function is
composed only by the third positive geometric decaying term and, in
consequence, the asymptotic hazard rate of this model is a constant. The mean
$\mu$ and standard deviation $\sigma$ of $P_{N=3}(n)$ are $9$ and
$27^{1/2}$ respectively, with an aperiodicity $\alpha=\sigma/\mu =
0.58$.
\\

\section*{Acknowledgements}
This work was supported by project FIS2005-06237 of the Spanish Ministry of Education and Science. A.F.P. thanks C. Sang\"{u}esa and J. As\'{i}n for insightful conversations. 
\\

\section*{References}
\bibliography{CModel_JPhysA}

\end{document}